# Formation and stability of reduced TiO$_x$ layers on anatase TiO$_2$(101): identification of a novel Ti$_2$O$_3$ phase

Xunhua Zhao∗, Sencer Selcuk, and Annabella Selloni∗

Department of Chemistry, Princeton University,

Princeton, NJ   08544, USA

We use density functional theory (DFT) calculations to investigate structural models consisting of anatase TiO$_2$(101) slabs covered by reduced overlayers formed by (101) crystallographic shear planes (CSPs). Ab initio thermodynamics supports the stability of these structures under a wide range of experimental conditions. The overlayers are found to have Ti$_2$O$_3$ stoichiometry with a crystal structure different from the known corundum-like Ti$_2$O$_3$ (here denoted α-Ti$_2$O$_3$) phase. DFT calculations predict this new "csp-Ti$_2$O$_3$" phase to be energetically close to α-Ti$_2$O$_3$ and to have also a similar band gap. These results suggest a possible role of the csp-Ti$_2$O$_3$ phase in the properties of black TiO$_2$, a promising photocatalytic material made of nanoparticles with a crystalline TiO$_2$ core and a highly reduced TiO$_x$ shell that is capable of absorbing the whole spectrum of visible light.

Nanomaterials of titanium dioxide ($TiO_2$) are of great interest in many fields, ranging from energy conversion and storage to coatings and biomedical applications [1,2]. In particular, recent research efforts have been focused on "black $TiO_2$", which consists of photocatalytically active core-shell nanoparticles (NPs) that absorb visible light much more efficiently than pristine $TiO_2$ [3,4]. It is widely accepted that the characteristic properties of black $TiO_2$ are due to structural changes in the outer shell of the NPs while the core maintains the crystal structure of conventional $TiO_2$ [3–9]. However, understanding of the detailed chemical composition and atomic structure of the outer shell has remained limited.

Independent of the synthetic procedure, a key feature of all black $TiO_2$ nanomaterials is that they are highly reduced (i.e. oxygen deficient), since they are prepared either by exposing the nanoparticles to a reducing atmosphere or by annealing the sample in vacuum to create oxygen vacancies [3,6]. As a prototypical reducible oxide, $TiO_2$ always contains a significant amount of oxygen vacancies ($V_O$s) [10]. With increasing concentration, $V_O$s are known to rearrange to give rise to crystallographic shear planes (CSPs) [11–13]. For example, the Magnéli phases, $Ti_nO_{2n-1}$, can be described as resulting from a regular arrangement of CSPs in oxygen-deficient rutile [12]. Interestingly, a recent study characterized the outer shell of black rutile $TiO_2$ nanoparticles as consisting of disordered $\alpha$-$Ti_2O_3$ [6]. Similarly, based on the $Ti^{3+}/Ti^{4+}$ ratio of Ti ions, another study determined the black shell of anatase $TiO_2$ nanotubes to be amorphous $Ti_4O_7$ [7]. Strips of CSPs have been reported also in anatase $TiO_2$(001) thin films epitaxially grown on $LaAlO_3$ [14], where they have been attributed to anatase-derived Magnéli-like phases $Ti_nO_{2n-1}$ ($n$=5,6,7).

In this work, we focus on anatase, the $TiO_2$ phase usually found in nanomaterials and most relevant in photocatalysis and photovoltaics [1,15,16], and use DFT calculations to explore the stability and properties of highly reduced $TiO_x$ overlayers formed by CSPs on or near its most stable and frequently exposed (101) surface [10,15,17]. While ordered, these CSP structures could indirectly provide information also on the importance of structural disorder on the properties of different black $TiO_2$ nanomaterials. Our results show that formation of a reduced shell on the anatase surface is thermodynamically favourable under a wide range of experimental conditions. This shell has $Ti_2O_3$ stoichiometry and its structure is not the well-known α-$Ti_2O_3$ phase (the mineral "tistarite" [18]), but a novel phase that has not yet been reported. DFT calculations with various exchange-correlation functionals (see the Supplemental Material [19], for computational details) show that this new phase – here denoted csp-$Ti_2O_3$ – is close in stability to α-$Ti_2O_3$ and has also a similar band gap, which is consistent with the absorption of black $TiO_2$. These findings suggest the possible presence of csp-$Ti_2O_3$ in the outer shell of black $TiO_2$ nanomaterials.

*Aggregation of $V_O$s and CSPs at the Anatase (101) Surface*- $V_O$s at slightly or moderately reduced anatase (101) have been found to reside not at the very surface, but one or a few layers below it [20,21]. In agreement with these findings, our calculations show that a subsurface $V_O$ (site 4 in Fig. 1) is energetically more favourable than one at a surface bridging oxygen (site 1 in Fig. 1) by 0.17 eV. Since $V_O$s at these sites are significantly more favourable than at any other site, we only consider such sites when studying the formation of multiple vacancies.

Also in agreement with previous studies, our calculations show a slight increase of the average formation energy per $V_O$ with increasing number of $V_O$s, indicating a weakly repulsive interaction between them (Table I). However, when a full layer of subsurface oxygen atoms is removed (Fig. 1), the formation energy ($E_{form}$) decreases sharply by 0.28 eV compared to the value of a single, most-stable $V_O$. This decrease is associated to a lateral shift of the layer above the one that is removed and the formation of a (101) crystallographic shear plane; in particular, the position of O4 is taken by O5, which becomes four-coordinated and shared by the first and second layer.

To determine the relative stabilities of CSPs and disordered $V_O$s, the contribution of the configurational entropy per oxygen vacancy, $S_{conf}$, must also be included [22]. Being perfectly ordered, a CSP has $S_{conf} = 0$, whereas separated $V_O$s can have a large number of different configurations, and therefore a large value of $S_{conf}$. As a simple estimate, we assume the $V_O$s to be non-interacting, and for any given stoichiometry $TiO_{2-x}$ we approximate $S_{conf}$ by $[k_B \ln(2/x)]$, where $2/x$ is the average number of available sites per $V_O$. We then calculate the minimum temperature, $T^*$, that is required for the disordered $V_O$s to be more stable than the CSP layer, i.e. $T^* \cdot S_{conf} \geqslant 0.28$ eV. In this way we obtain $T^*$= 610, 425, 326 and 265 K at x= $10^{-2}$, $10^{-3}$, $10^{-4}$, and $10^{-5}$, respectively, indicating that CSPs are stable at room temperature for x > $10^{-4}$. More detailed analysis of the contribution of $S_{conf}$, presented in the Supplemental Material [19], also supports this conclusion.

We further considered structures with several contiguous CSPs (denoted 2CSP, 3CSP, etc.) obtained by sequentially removing full layers of oxygen atoms. The average $V_O$ formation energy decreases with increasing number of CSPs, slowly reaching a saturation value of ~ 3.40 eV (Table I). We also investigated whether CSPs prefer to be contiguous or separated from each other, and whether they prefer to form close to the surface or in the central bulk-like region of the anatase slab. Computed formation energies for different models clearly indicate that CSPs prefer to aggregate and remain close to the surface (Fig. S3 [19]).

The densities of states (DOS) for slabs with different numbers of CSPs are shown in Fig. 2. When there is one CSP, the gap states are similar to those from isolated $V_O$s, with some additional broadening. Two CSPs give rise to two slightly separated sets of gap states. The layer-resolved DOS reveals that the lower set is contributed mainly by Ti 3d orbitals in the second layer (Figs. S5-S7 [19]). As the number of CSPs increases to 3, the distribution of gap states becomes even broader due to the presence of many nonequivalent Ti atoms at the interface and in the central layers. As a result, the energy gap between the occupied Ti3d states and the empty states in the conduction band is only 0.2 eV. However, when the number of CSPs reaches 4 or more, the energy gap increases again to 0.66, 0.67 and 1.18 eV for $n$=4,5, and 6 respectively. This increase can be attributed to the formation of a new insulating phase, csp-$Ti_2O_3$, that is discussed below.

*A new csp-$Ti_2O_3$ crystal phase* - As the number of CSPs increases, the presence of a crystalline structure in the central region of the CSPs becomes evident. This structure, denoted csp-$Ti_2O_3$, has $Ti_2O_3$ stoichiometry, but is different from the usual α-$Ti_2O_3$ phase. csp-$Ti_2O_3$ is orthorombic (Fig. 3) with space group I*mmm* (no. 71), and can be considered as the limiting structure of a series of Magnéli $Ti_nO_{2n-1}$ phases (Fig. S9 [19]). The primitive unit cells of bulk csp-$Ti_2O_3$, α-$Ti_2O_3$ and anatase $TiO_2$ are shown in Fig. 3, while computed structural parameters are given in Tables S1 and S3 [19]. csp-$Ti_2O_3$ has only one type of Ti site but two inequivalent oxygen sites, both four-fold coordinated: one (O1) is at the center of a distorted tetrahedron and the other (O2) forms four O-Ti bonds in a plane. The Ti-O bonds are elongated compared to anatase, while Ti-Ti distances − one shorter than the other two in the unit cell − suggest the formation of Ti-Ti dimers as in α-$Ti_2O_3$.

It is interesting to compare the relative stabilities of csp-$Ti_2O_3$ and α-$Ti_2O_3$. To address this question we performed DFT+$U$ [23] calculations with the Perdew-Burke-Ernzerhof (PBE) [24] functional and on-site Coulomb repulsion $U$ on the Ti 3d orbitals, considering 10 different $U$ values in the range between 0 and 5 eV, which includes the values typically used for $Ti^{4+}$ and $Ti^{3+}$ ions in titanium oxides [25,26]. As a further check, we also used the hybrid HSE06 functional [27], which is known to work well for many oxides including $TiO_2$ [28]. Experimentally, α-$Ti_2O_3$ is observed to be a small band gap insulator with no long-range magnetic ordering at room temperature.[29,30] We found that both α-Ti$_2$O$_3$ and csp-Ti$_2$O$_3$ are metallic and either non-magnetic or with a small absolute magnetization (Table S5 [19]) for $U \leq 1.5$ eV, whereas they are insulating and antiferromagnetic (AFM) for larger $U$ and with HSE06 as well. For comparison, α-Ti$_2$O$_3$ was found to be AFM and insulating (with band gap E$_g$ = 0.59 eV) in a recent study using self-consistent hybrid functionals [31], and nonmagnetic and insulating (with E$_g$ = 0.22-0.57 eV) in studies employing screened exchange and modified (LDA-based) HSE06 hybrid functionals [32,33].

Computed values of the total energy difference between α-Ti$_2$O$_3$ and csp-Ti$_2$O$_3$ are reported in Fig. 4 (see also Table S4 [19]). Two different choices of lattice parameters were considered. As experimental lattice parameters are not available for csp-Ti$_2$O$_3$, we used the lattice parameters optimized with HSE06, which are very close to the experimental ones in the case of α-Ti$_2$O$_3$ and anatase TiO$_2$ (Table S1 [19]). We further computed PBE+$U$ energy differences using the lattice parameters also optimized with PBE+$U$, which tend to deviate significantly from experiment, especially for large $U$ [25]. From Fig. 4, it is evident that the results obtained with the two choices of geometries are quite different, as PBE+ $U$ predicts csp-Ti$_2$O$_3$ to be more (less) stable than α-Ti$_2$O$_3$ for $U > 2$ ($U > 1.5$) when HSE06 (PBE+$U$) optimized geometries are used. Interestingly, the small range of $U$ values, 1.5 $\leq U \leq$ 2 eV, for which the relative stability of the two phases is independent of the choice of the lattice constant is located around the transition between metallic and insulating behaviour, where the computed band gap for α-Ti$_2$O$_3$ (on the insulating side) is close to the experimental value (see Table S4 [19] and below) and the reduction energy of TiO$_2$ to Ti$_2$O$_3$ is also in good agreement with experiment [25] (see also Fig. S1 [19]). Here, csp-Ti$_2$O$_3$ is predicted to be less stable than α-Ti$_2$O$_3$, as one would expect, but only by a small amount.

The electronic DOS of csp-$Ti_2O_3$ and α-$Ti_2O_3$ computed at the PBE+$U$ ($U$=1.75 and 3 eV), and HSE06 levels are shown in Fig. 5 (see also Figs. S10 and S11 [19]). Results for the two phases are similar despite their significant structural differences. In particular, $U$=1.75 eV gives $E_g$ = 0.30 eV for α-$Ti_2O_3$, close to experimental value of 0.1-0.2 eV [34,35], while HSE06 predicts much larger band gaps of 1.90 eV and 1.82 eV for α-$Ti_2O_3$, and csp-$Ti_2O_3$, respectively. For comparison, Table S4 [19] reports also the computed band gap of anatase, which is in the range 2.2-3.7 eV with PBE+$U$ and HSE06. Independent of the approach, all methods predict that the band gap of csp-$Ti_2O_3$ is close to that of α-$Ti_2O_3$ and about 1-2 eV smaller than the band gap of anatase $TiO_2$.

*Stability Diagrams* - To evaluate the relative stabilities of the csp-$Ti_2O_3$/$TiO_2$ structures with different degrees of reduction, Gibbs free energies of formation are calculated in either $O_2$ or $H_2$/$H_2O$ atmosphere [36,37]. We used a 7-layer pristine anatase $TiO_2$(101) slab as the reference, and considered the slab models listed in Table I as well as bulk csp-$Ti_2O_3$. Based on the results in Fig. 6(a), formation of CSPs in a $TiO_2$ slab or formation of bulk csp-$Ti_2O_3$ from bulk $TiO_2$ (see inset) in a pure $O_2$ atmosphere requires the $O_2$ pressure to be well below the pressures that can be achieved in ultra-high vacuum experiments.

Alternatively, one of the common procedures for preparing black $TiO_2$ is by reducing anatase powders under $H_2$ atmosphere. This reduction proceeds according to $H_2$+$O_s$ → $H_2O$ + $V_O$, where $O_s$ is a surface oxygen atom [4,38,39]. We thus evaluated the thermodynamics of forming $n$CPSs@$TiO_2$ and bulk csp-$Ti_2O_3$ from anatase $TiO_2$ in the presence of a mixed $H_2$/$H_2O$ environment, where the reduction of $TiO_2$ is driven by the formation of water. Phase stability diagrams as a function of the H chemical potential ($\Delta\mu_H$) with $H_2O$ at high (ambient condition) and low partial pressures are shown in Fig. 6(b) and Fig. S8 [19]. In the mixed $H_2$/$H_2O$ atmosphere, formation of the reduced structures becomes favorable at values of the chemical potentials that are well accessible experimentally. Although this thermodynamic analysis only predicts the preferred state at infinitely long time, without taking into account the kinetics that could actually play a critical role, e.g. in determining the thickness of the reduced layer, it is evident that the combination of oxygen from $TiO_2$ with hydrogen in the surrounding environment provides a strong driving force toward the formation of CSPs and eventually of the csp-$Ti_2O_3$ phase on the anatase surface.

In summary, based on DFT calculations, we have identified a new metastable csp-$Ti_2O_3$ phase that results from aggregation of CSPs on anatase $TiO_2$(101) and is close in stability to α-$Ti_2O_3$. Structures consisting of a csp-$Ti_2O_3$ overlayer on crystalline anatase are thermodynamically stable in reducing atmosphere and, depending on the thickness of the overlayer, have energy gaps in the range 0.2 ~ 1.2 eV, consistent with the photoabsorption of black $TiO_2$. Besides representing viable models for black $TiO_2$ nanomaterials, such structures might be relevant for understanding recent experimental observations on anatase (101) chemically doped with oxygen vacancies [40], and for describing the reduced surface region of nanomaterials of other metal oxides as well [41,42].

This work was supported by DoE-BES, Division of Chemical Sciences, Geosciences and Biosciences under Award DE-FG02-12ER16286. We used resources of the National Energy Research Scientific Computing Center (DoE Contract No. DE-AC02-05CH11231). We also acknowledge use of the TIGRESS high performance computer center at Princeton University.

∗ x.zhao@princeton.edu

∗ aselloni@princeton.edu

TABLE I: Formation energy per oxygen vacancy $E_{\text{form}}$ (eV), computed using PBE+$U$ with $U$=3.0 eV, for different model systems: one and two vacancies in a 7 layer anatase (101) slab, $n$CSP@(7-$n$)TiO$_2$ ($n$ = 1- 6) with $n$ CSPs in a 7 layer (101) slab, and bulk csp-Ti$_2$O$_3$ (last column). For the latter, $E_{\text{form}}$ is computed from the reaction 2TiO$_2$ → Ti$_2$O$_3$ + ½ O$_2$.

| Model | Vo | 2Vo | 1CSP | 2CSP | 3CSP | 4CSP | 5CSP | 6CSP | csp-Ti$_2$O$_3$ |
|---|---|---|---|---|---|---|---|---|---|
| $E_{\text{form}}$ | 3.92 | 3.94 | 3.65 | 3.51 | 3.46 | 3.44 | 3.42 | 3.44 | 3.39 |

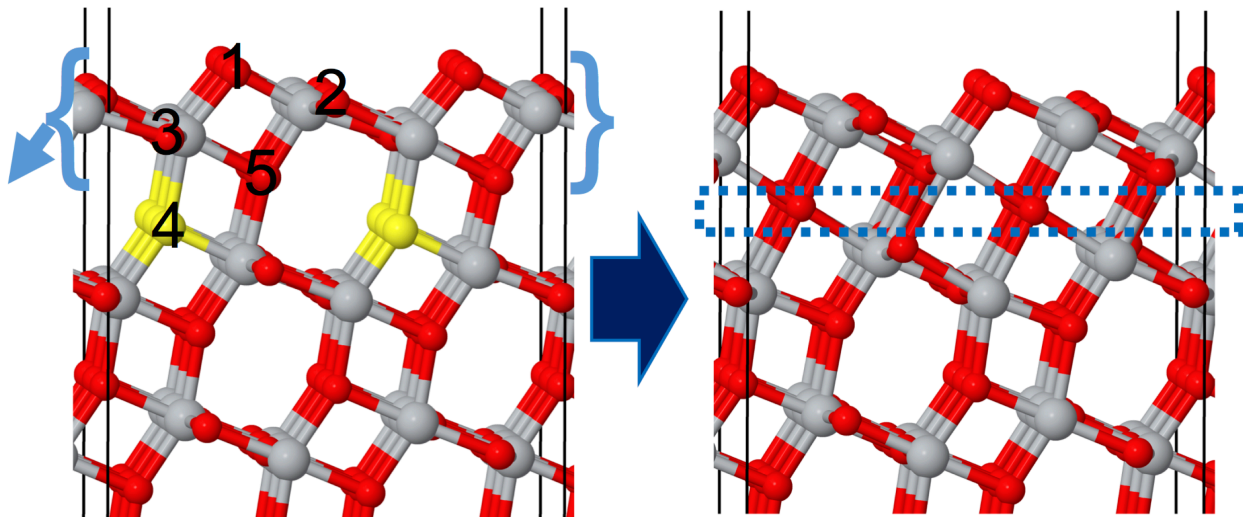

FIG. 1: Formation of a CSP parallel to the (101) surface (schematic). O atoms removed upon CSP formation are yellow, all other O atoms are red, Ti atoms are grey. Five non-equivalent oxygen atoms are indicated. The part within bracket is the top layer of $TiO_2$; the arrow shows the direction of shear.

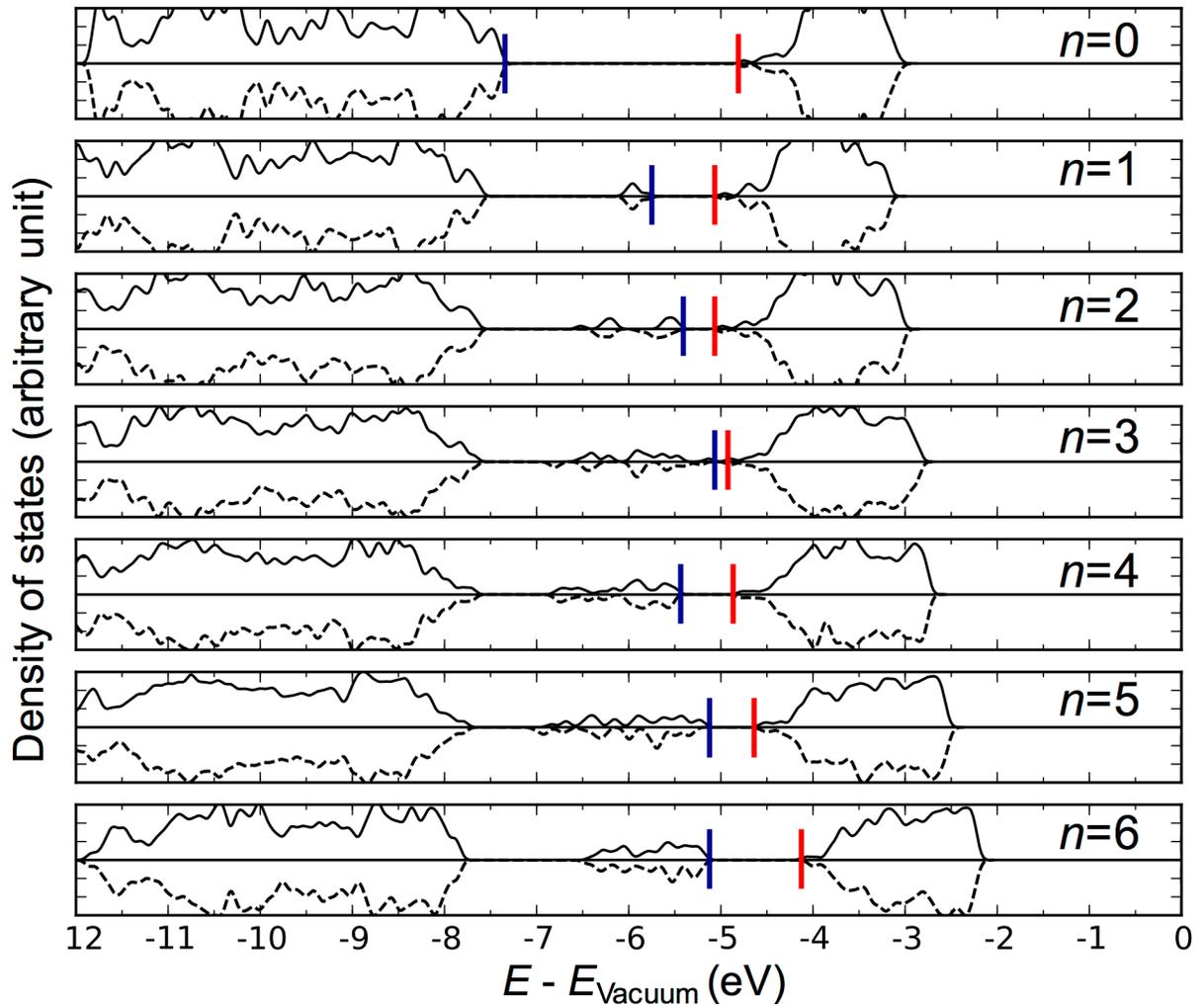

FIG. 2: Electronic densities of states of $n$CSP@$(7-n)$TiO$_2$ ($n$=0-6), computed using PBE+$U$ with $U = 3.0$ eV. Solid and dashed lines refer to spin up and spin down states. Blue and red bars indicate valence and conduction band edges, respectively. The zero of energy is set at the vacuum level (see Fig. S4 [19])

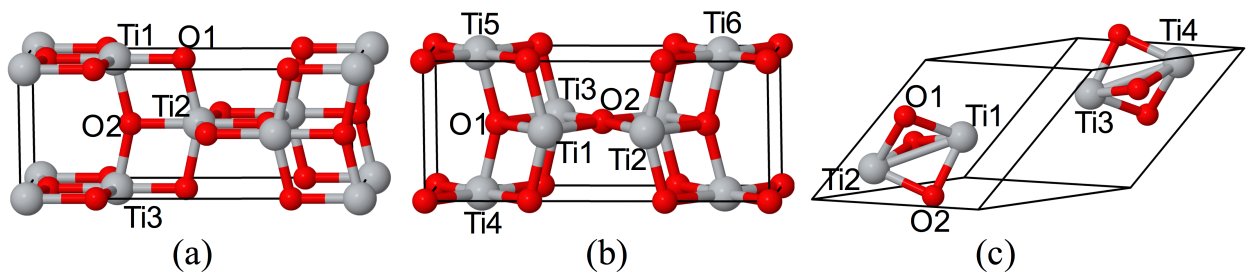

FIG. 3: Primitive unit cells of (a) anatase $TiO_2$, (b) csp-$Ti_2O_3$ and (c) α-$Ti_2O_3$. Oxygen atoms are red, titanium atoms are grey.

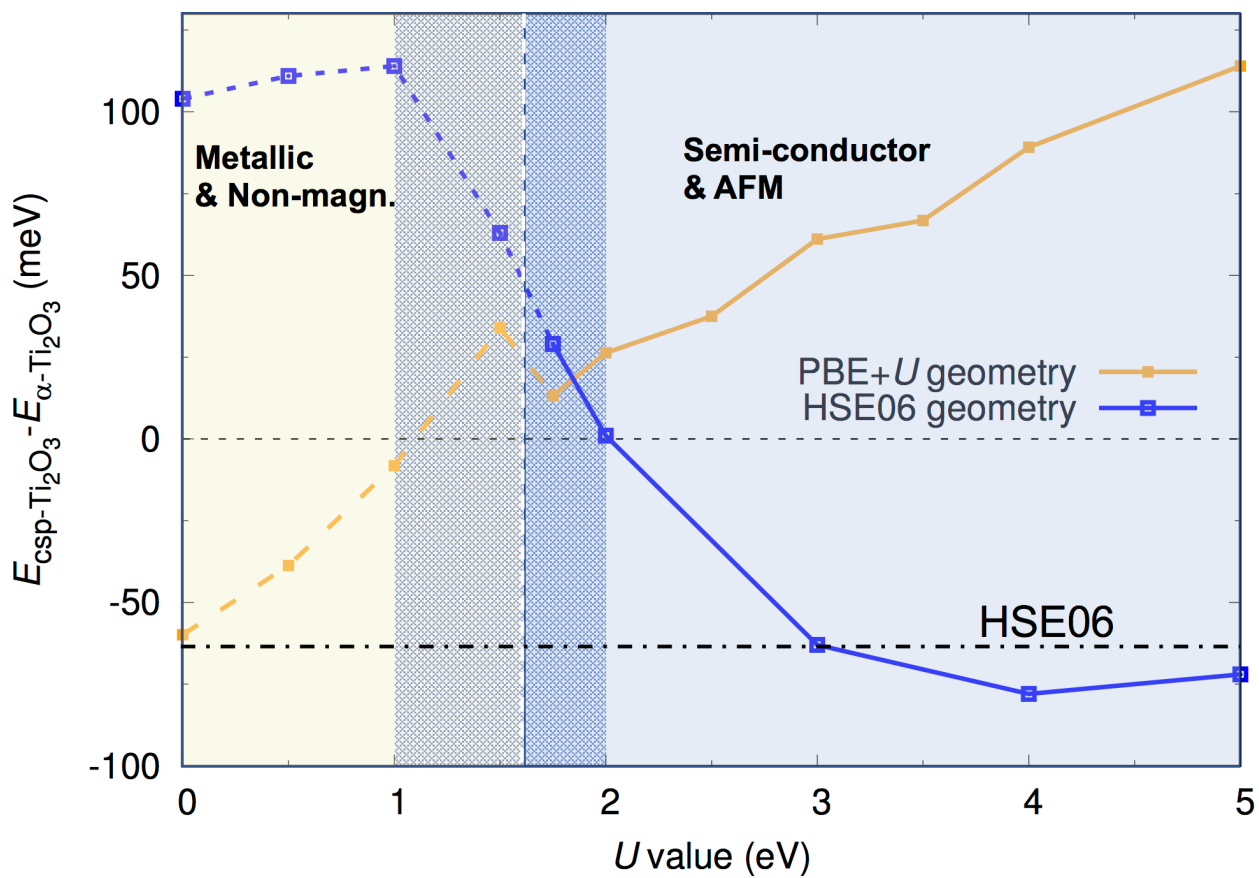

FIG. 4 Total energy difference per formula unit between csp-$Ti_2O_3$ and α-$Ti_2O_3$, computed using PBE+$U$ with different $U$ values, and two different choices of lattice parameters. In the region $1 <$ $U < 2$ eV (darker shading), the magnetic moment of $Ti^{3+}$ ions increases from 0 to nearly 1 $μ_B$ (Table S5 [19]). The horizontal dashed-dotted line indicates HSE06 results.

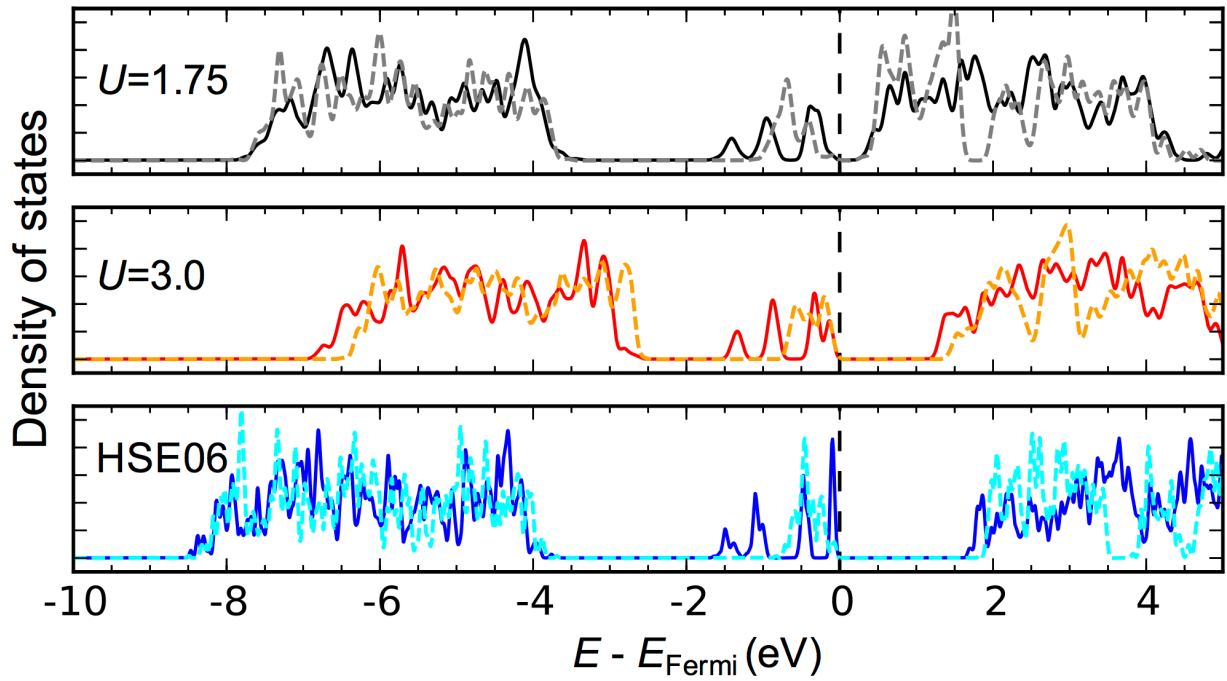

FIG. 5: Density of states of bulk csp-$Ti_2O_3$ (solid lines) and α-$Ti_2O_3$ (dashed lines), computed using PBE+$U$ ($U$=1.75 and 3.0 eV) and hybrid HSE06. The Fermi level is at the top of the occupied states. DOS curves for spin-up and spin-down states are identical.

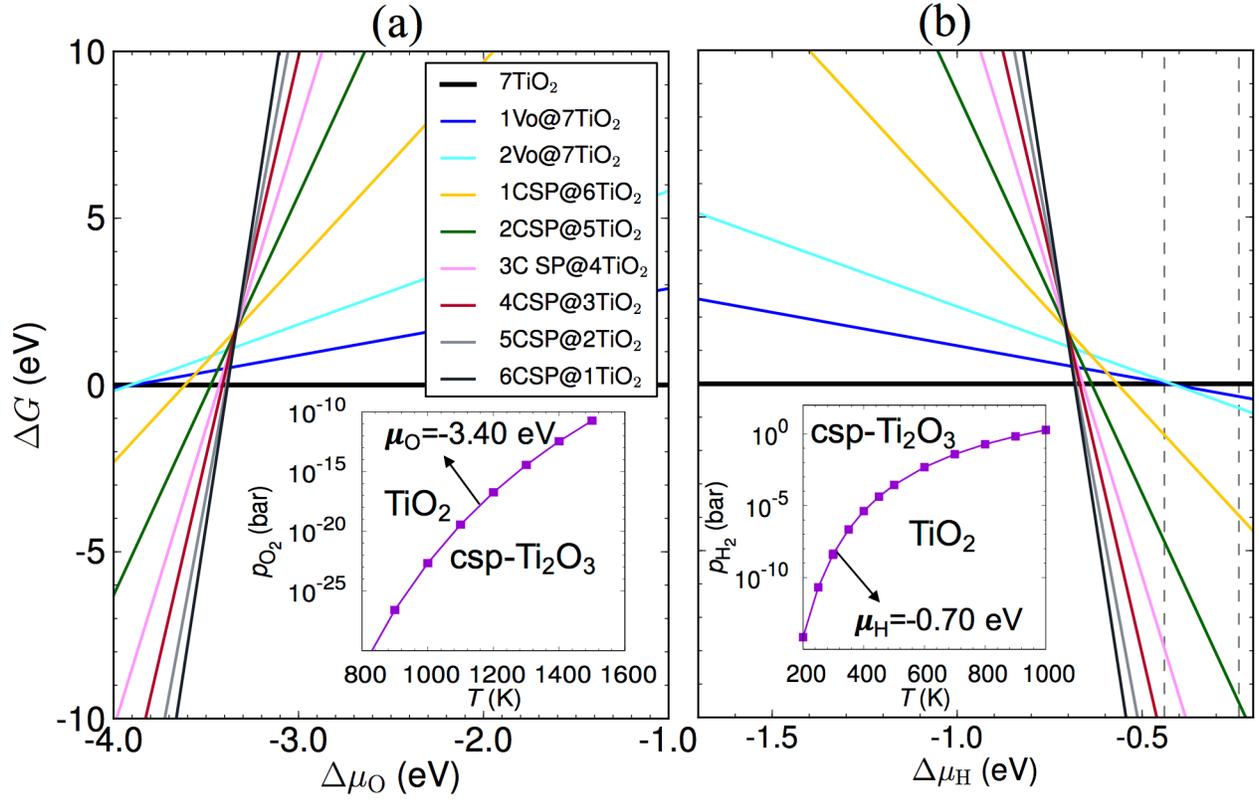

FIG. 6: Formation free energy ($\Delta G$) for the structures in Table I as function of: (a) O chemical potential (relative to 1/2 $O_2$), and (b) H chemical potential (relative to 1/2 $H_2$), with $H_2O$ at $T = 800$ K and $p_{H2O} = 10^{-2}$ bar; the two dashed vertical lines mark the region of H chemical potential typical of black $TiO_2$ synthesis. The insets show the phase diagram for bulk csp-$Ti_2O_3$ and $TiO_2$ as a function of temperature ($T$) and $O_2$(a) or $H_2$(b) pressure. $\Delta G$ for $nV_O@7TiO_2$ ($n=1,2$) do not include the configurational entropy discussed in the text.